\documentclass[a4paper]{jpconf}
\usepackage{graphicx}

\usepackage{hyperref}
\usepackage{xspace}
\usepackage{lineno}

\def \pao      {Pierre Auger Observatory\xspace}
\def \degree   {$^{\circ}$\xspace}

\begin{document}
\title{Atmospheric considerations for CTA site search using global models}
\author{Karim Louedec$^{1}$ and Martin Will$^{2}$}
\address{$^1$ Laboratoire de Physique Subatomique et de Cosmologie (LPSC), UJF-INPG, CNRS/IN2P3, Grenoble, France.}
\address{$^2$ Karlsruher Institut f\"ur Technologie, Institut f\"ur Kernphysik, Karlsruhe, Germany.}
\ead{karim.louedec@lpsc.in2p3.fr, martin.will@kit.edu}

\begin{abstract}
  The Cherenkov Telescope Array (CTA) will be the next high-energy gamma-ray
  observatory. Selection of the sites, one in each hemisphere, is not obvious
  since several factors have to be taken into account. Among them, and probably
  the most crucial, are the atmospheric conditions. Indeed, CTA will use the
  atmosphere as a giant calorimeter, i.\,e.\ as part of the detector. The
  Southern Hemisphere presents mainly four candidate sites: one in Namibia, one
  in Chile and two in Argentina. Using atmospheric tools already validated in
  other air shower experiments, the purpose of this work is to complete studies
  aiming to choose the site with the best quality for the atmosphere. Three
  strong requirements are checked: the cloud cover and the frequency of clear
  skies, the wind speed and the backward trajectories of air masses travelling
  above the sites and directly linked to the aerosol concentrations. It was
  found, that the Namibian site is favoured, and one site in Argentina is
  clearly not suited. Atmospheric measurements at these sites will be performed
  in the coming months and will help with the selection of a CTA site.
\end{abstract}


The ground-based observation of high-energetic photons uses the atmosphere as a
giant calorimeter. Photons interact in the upper atmosphere with molecules in
the air and produce air showers, cascades of billions of secondary charged
particles. These particles travel through the air faster than the speed of light
in the medium, creating Cherenkov light in the process. The amount of Cherenkov
light emitted per metre can be used to perform a calorimetric measurement of the
shower energy. The detectors of the Cherenkov Telescope Array (CTA) will be
spaced several metres apart on an area of a few square kilometres to measure the
same air shower from different angles~\cite{CTA_consortium}. This enables a
better angular reconstruction of the shower axis and improves the estimate of
the shower energy. To reconstruct the energy accurately, the atmospheric
conditions have to be known very well. The Cherenkov light yield depends on the
refractive index, which in turn depends on temperature and pressure. Also, the
photons are attenuated on their way to the detector by Rayleigh and aerosol
scattering.

Four possible locations for the CTA site on the Southern Hemisphere are
evaluated. The site on the African continent is located on the Khomas Highland
Plateau in the Khomas region of Namibia. The coordinates are 23\degree~{\bf S},
16\degree~{\bf E}. This point is close to the site of the H.E.S.S.~\cite{HESS}
gamma ray observatory. Three sites are evaluated in South America, La Silla in
the Coquimbo region of Chile (29\degree~{\bf S}, 70\degree~{\bf W}, close to the
La Silla Observatory), El Leoncito National Park in the San Juan province of
Argentina (32\degree~{\bf S}, 69\degree~{\bf W}, close to the Leoncito
Astronomical Complex) and San Antonio de Los Cobres in the Salta province of
Argentina (24\degree~{\bf S}, 66\degree~{\bf W}).

\section{The Global Data Assimilation System and its application to the candidate sites}

Height-dependent profiles of the state variables of the atmosphere can be
measured by weather balloons. An extensive balloon program is very costly and
requires trained personnel and special equipment. Balloon launches are performed
at several places around the world on a regular basis, mostly on airports to
monitor the local conditions for air travel. These data are available online and
can be used in the reconstruction of air showers. The temporal resolution is on
the order of 12~hours. CTA will most likely be constructed on an arid high
plateau with mostly clear atmospheric conditions and low cloud cover. In such an
environment, the atmospheric state variables tend not to change too much in the
course of one night. Therefore, the resolution of the airport balloon data is
sufficient. CTA cannot be constructed too close to cities due to air and light
pollution. This also means the site will be far away from any airports, and the
validity of atmospheric data might not be given over too large distances. In
these cases, data from numerical weather prediction can be used.

The Global Data Assimilation System (GDAS) is based on forecast data from
numerical weather prediction which is supplemented with real-time atmospheric
measurements around the globe. This includes weather stations, balloon data and
satellite measurements. The GDAS data are modelled on a one degree
latitude-longitude grid ($360$\degree $\times$ $180$\degree), with a temporal
resolution of three hours. GDAS provides 23~pressure levels, i.\,e.\
23~different altitudes, from $1000$~hPa (more or less sea level) to $20$\,hPa
(around $26$\,km). Lateral homogeneity of the atmospheric variables across
several kilometres can be assumed on arid high plateaus. The validity of GDAS
data in such conditions was previously studied for the site of the \pao in
Argentina. The agreement with ground-based weather stations and on-site
meteorological radiosonde launches is very good for the state variables of the
atmosphere~\cite{GDASpaper}, as well as the wind~\cite{KarimECRS}.

\subsection{Validity of the GDAS model at the Namibian site}

For the four aforementioned sites, GDAS data were extracted for 2009 and 2010
from weekly data files available through the Real-time Environmental
Applications and Display sYstem (READY) website~\cite{websiteNOAA}. For the
Namibian site, the GDAS data were compared to radiosonde launches performed
twice daily at the Windhoek Hosea Kutako International Airport, about 150\,km
away from the extracted GDAS point. The difference was found to be about 1--2\,K
in temperature, 2\,hPa in pressure at all altitudes. The water vapour pressure
differs by about 1\,hPa in the lowest 3\,km above ground. These differences are
expected across distances of more than 100\,km. It has to be mentioned, that the
Windhoek airport data is used in the assimilation process of the GDAS data, so
both datasets are not independent.

\begin{figure*}[!t]
  \centerline{
    {\includegraphics [scale=0.42] {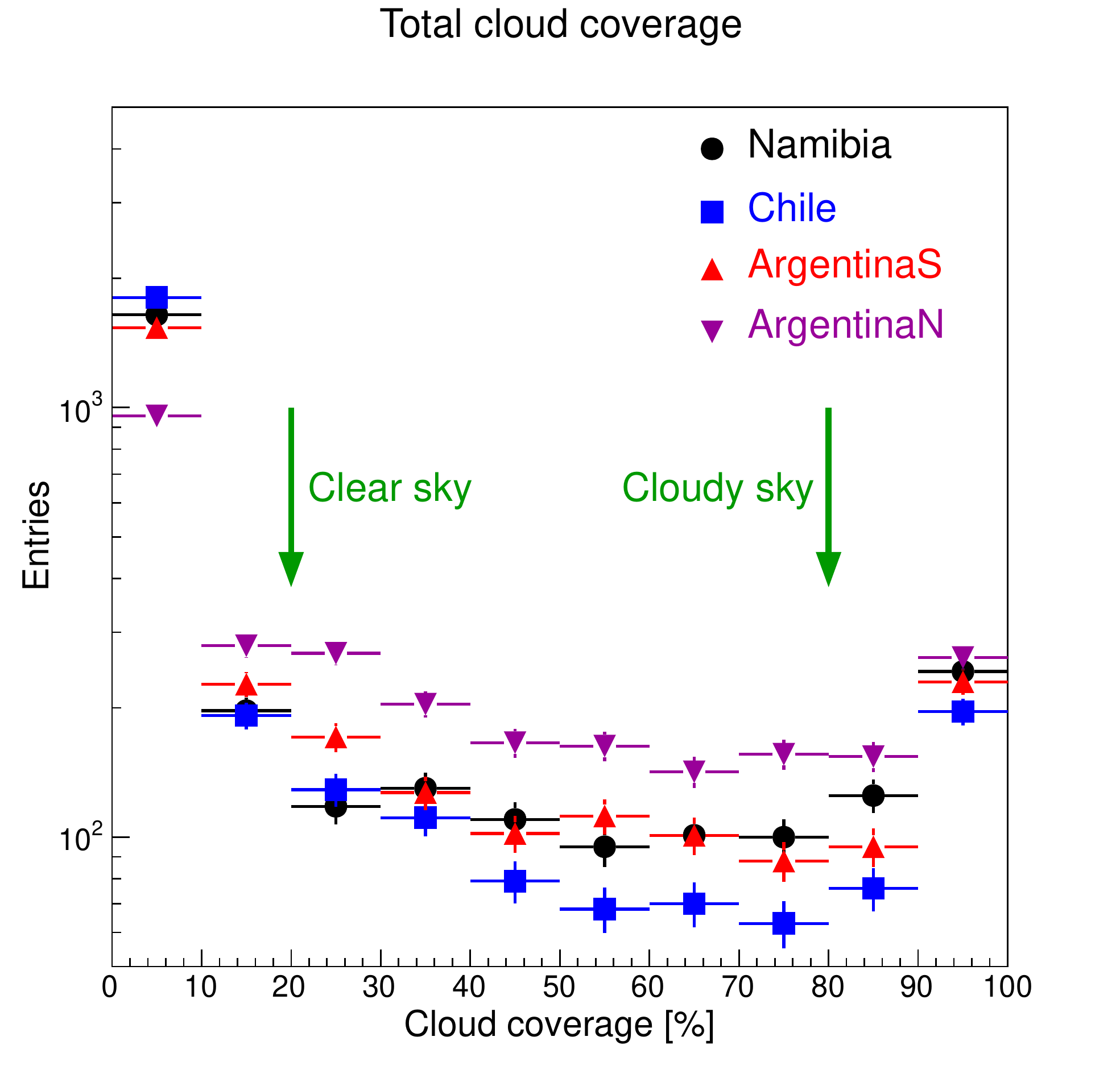}}
    \hfill
    {\includegraphics [scale=0.42] {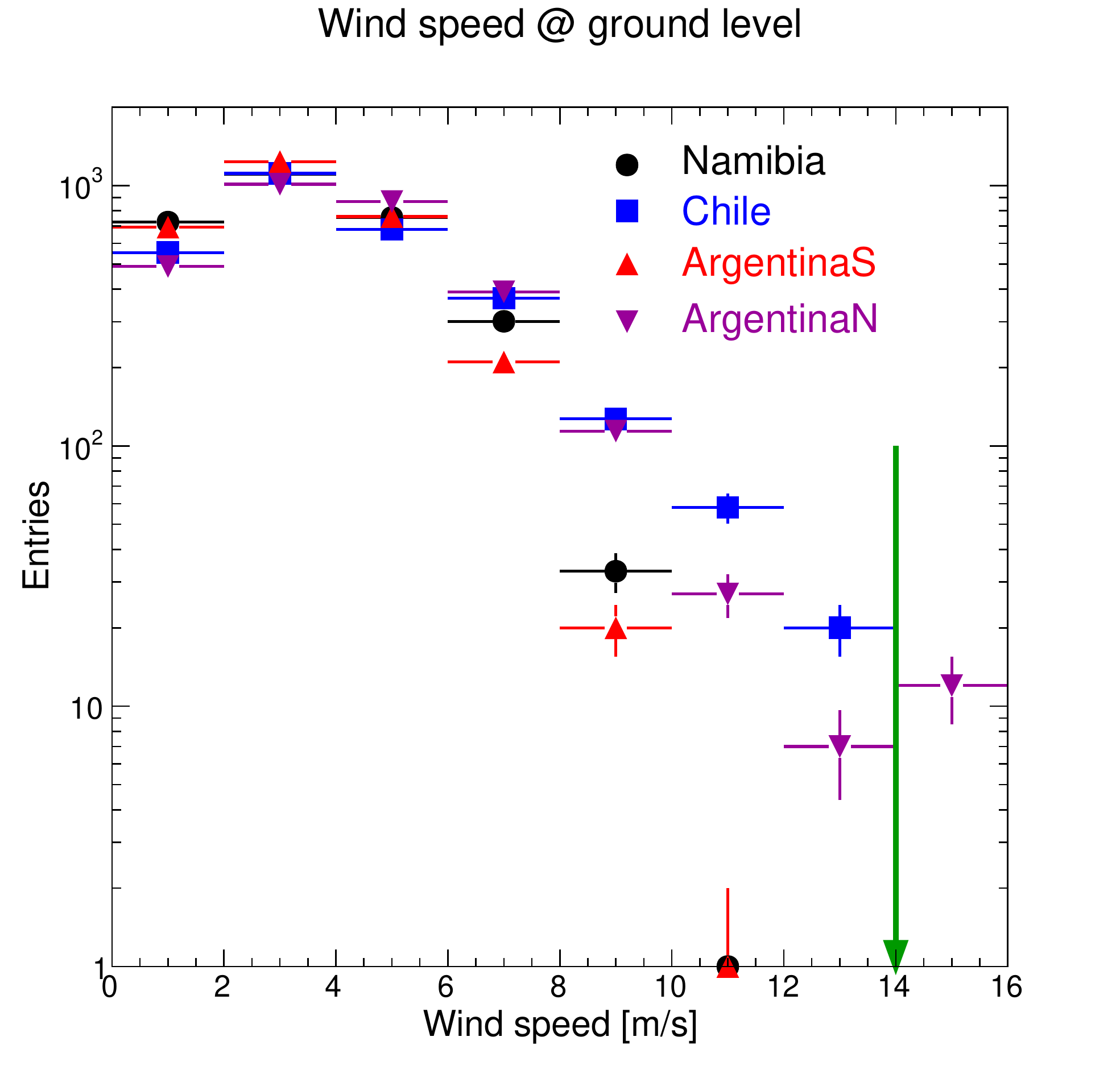}}
  }
  \caption{\label{fig:GDAS_cloud_wind}
    {\bf Cloud cover (left) and wind speed (right) distributions for the four
    sites in 2010 using the GDAS model.} An estimation of each value is provided
    every 3 hours. ArgentinaN and ArgentinaS correspond to the northern and
    southern sites, respectively. The arrow drawn for the wind speed
    distributions represent the speed where no observations are performed.
  }
\end{figure*}

\subsection{Estimate of cloud cover and wind speed at ground using the GDAS model}

Two of the most important requirements related to weather conditions are the
cloud cover and the wind speed at ground level. The fraction of clear nights has
to be as high as possible, typically a cloud cover lower than $20\%$ during more
than $80\%$ of the CTA data-taking. Wind speed affects also operation
efficiency. Wind speeds above $8~$m/s may impact the quality of the observations
and at speeds above $14~$m/s no measurements should be
performed~\cite{CTA_site_ICRC}. In Figure~\ref{fig:GDAS_cloud_wind}, left panel,
the distribution of cloud cover for the four sites in 2010 is shown. The two
Argentinian sites present the lowest numbers of clear sky configurations whereas
the Chilean site seems the best one. In the right panel of
Figure~\ref{fig:GDAS_cloud_wind}, wind speed distributions are plotted for the
year 2010. The Namibian and the southern Argentinian site present the lowest
values, with no values beyond the operational limit of $14~$m/s. In conclusion,
combining the two main requirements listed previously, the site based in Namibia
seems the best one to install the future CTA for the Southern Hemisphere. In
Figure~\ref{fig:Namibia_GDAS_HYSPLIT}, left panel, the monthly distribution of
clear ($\leq 20\%$) and cloudy skies ($\geq 80\%$) in 2010 is shown for the
Namibian site. The sky is almost completely clear during the entire summer (June
to August). Compared to the other sites studied previously, Namibia presents the
highest number of hours of clear sky.

\section{Origin of air masses in Namibia using the air-modelling program HYSPLIT}

Another important part before deciding where to install the telescope array is
to know the origins of the air masses above the telescopes during data-taking. A
study done by the Pierre Auger Collaboration has shown that air masses coming
more directly from the oceans without travelling over land for too long had a
lower aerosol concentration when they were passing above the
array~\cite{KarimECRS}. This conclusion leads to claim that sites close to the
coasts should have a clearer sky in terms of aerosol concentration. Aerosols are
the most variable term affecting the photon propagation in the atmosphere. It
depends mostly on the local environment and long-range transportation.

\begin{figure*}[!t]
  \centerline{
    {\includegraphics [scale=0.42] {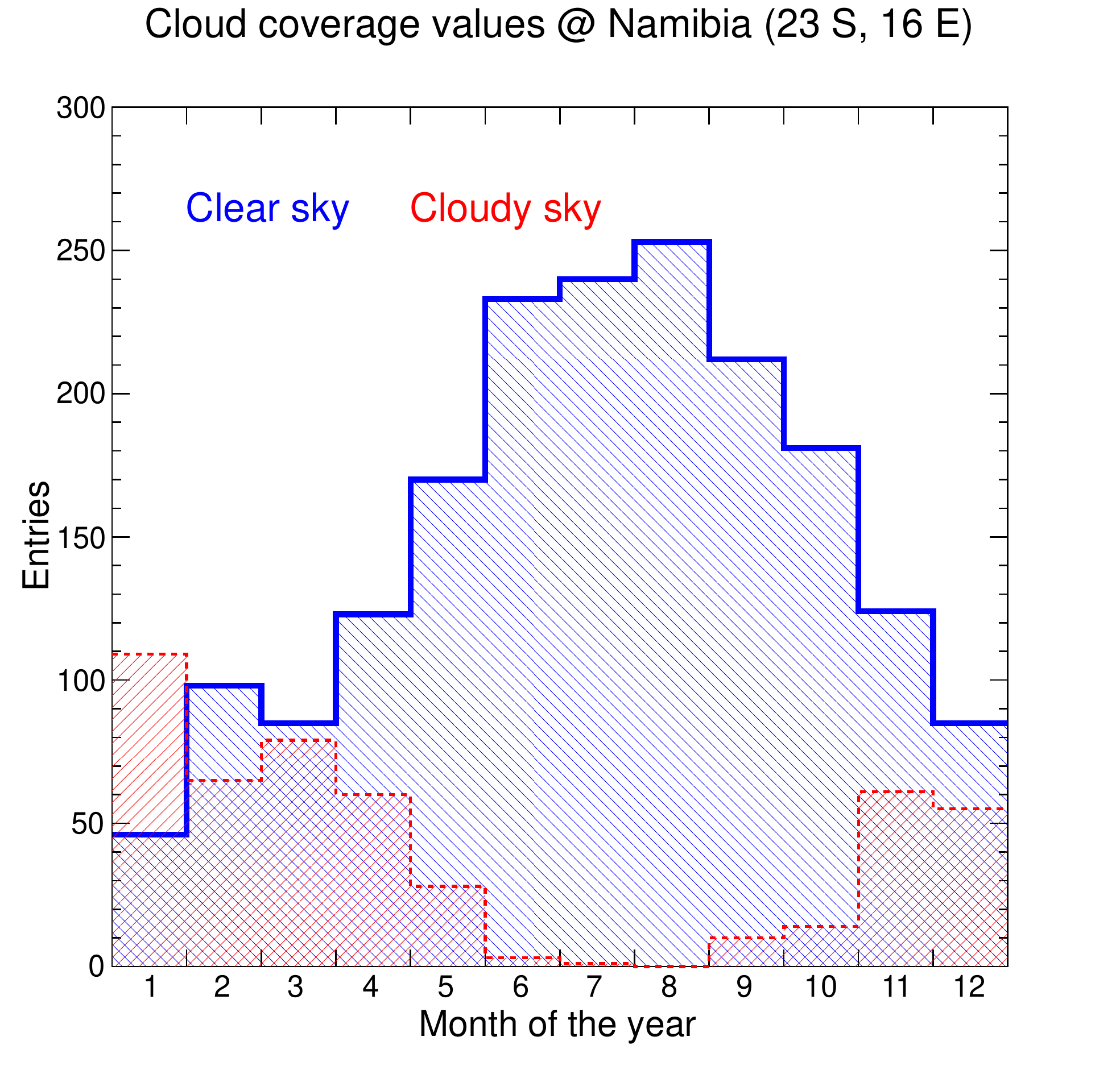}}
    \hfill
    {\includegraphics [scale=0.42] {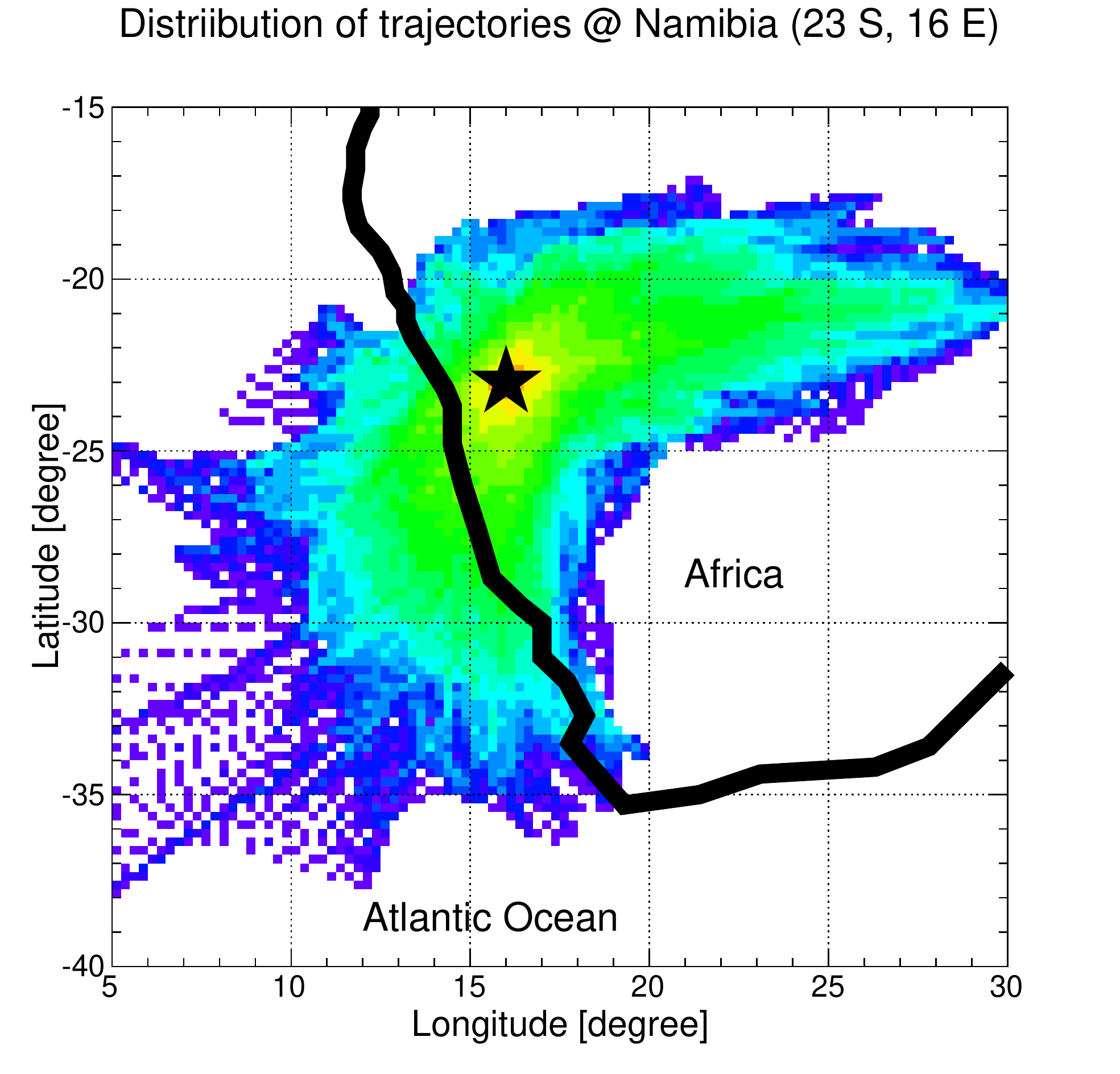}}
  }
  \caption{\label{fig:Namibia_GDAS_HYSPLIT}
    {\bf Atmospheric considerations for the Namibian site.} On the left, monthly
    distribution of clear skies ($\leq 20\%$, continuous line) and cloudy skies
    ($\geq 80\%$, dotted line) for the Namibian site using the GDAS model. On
    the right, distribution of backward trajectories of air masses in 2010 using
    HYSPLIT.  The black star and the black line represent the Namibian site
    location and the South-West African coast, respectively.
  }
\end{figure*}

Different air models have been developed to study air mass relationships between
two regions. Among them, the {\it HYbrid Single-Particle Lagrangian Integrated
Trajectory} model, or HYSPLIT~\cite{HYSPLIT_1,HYSPLIT_2}, is a commonly used
air-modelling program in atmospheric sciences that can calculate air mass
displacements from one region to another. The HYSPLIT model, developed by the
Air Resources Laboratory, NOAA (National Oceanic and Atmospheric
Administration)~\cite{websiteNOAA}, computes simple trajectories to complex
dispersion and deposition simulations using either puff or particle approaches
with a Lagrangian framework. A backward trajectory is computed over 48 hours
every hour, throughout the year 2010. The start altitude is fixed at 500\,m
above ground level. In Figure~\ref{fig:Namibia_GDAS_HYSPLIT}, right panel, the
distribution of the backward trajectories in 2010 is displayed for the Namibian
site. Air masses come mainly from the West, after having just a few hours above
continental areas. Following the conclusion presented in~\cite{KarimECRS},
aerosol component should be optimised for such telescope observations. Lidar
measurements performed at the H.E.S.S.\ site could confirm these
observations~\cite{HESS_LIDAR}.

\section{Conclusion}

CTA will be the next generation of imaging atmospheric Cherenkov telescopes.
Using atmospheric global models, this work presents the atmospheric conditions
for the main candidates for the experiment in the Southern Hemisphere. Regarding
requirements on wind speed at ground or cloud cover, the site close to the
H.E.S.S.\ site in Namibia seems the best location. Backward trajectories of air
masses passing through the Namibian site confirm this conclusion. Weather
stations built by the CTA Consortium, the so-called ATMOSCOPES, can be used to
verify these conclusions from global models.

\ack
The authors gratefully acknowledge the NOAA Air Resources Laboratory (ARL) for
the provision of the HYSPLIT transport and dispersion model, the Global Data
Assimilation System and the READY website.


\end{document}